\pdfoutput = 1
\documentclass[journal]{IEEEtran}
\usepackage{tikz}
\usetikzlibrary{shapes.geometric}
\usetikzlibrary{shapes,arrows}
\usetikzlibrary{fit,positioning,arrows}
\usepackage{float,adjustbox}
\usepackage{csquotes}
\usetikzlibrary{patterns}
\usepackage{cite}
\usepackage{amsmath,amssymb,amsfonts}
\usepackage{array}
\usepackage{algorithmic}
\usepackage{subfig}
\usepackage{graphicx}
\usepackage{textcomp}
\usepackage{xcolor}
\usepackage{physics}
\usepackage{multirow}

\usepackage{mathrsfs,bm,bbm}
\usepackage{csquotes}
\allowdisplaybreaks

\newcommand{\BlackBox}{\rule{1.5ex}{1.5ex}}  
\newenvironment{proof}{\par\noindent{\bf Proof\
}}{\hfill\BlackBox}
\newtheorem{theorem}{\bf{Theorem}}

\begin{document}

\title{
Stochastic Dynamic Network Utility Maximization with Application to Disaster Response
}

\author{
Anna Scaglione$^{(1)}$, Nurullah Karako\c{c}$^{(2)}$\\
$~^{(1)}$ 
Cornell University\\
$~^{(2)}$ 
Arizona State University}

\maketitle

\begin{abstract}
In this paper, we are interested in solving Network Utility Maximization (NUM) problems whose underlying local utilities and constraints depend on a complex stochastic dynamic environment. 
While the general model applies broadly, this work is motivated by resource sharing during disasters concurrently occurring in multiple areas. In such situations, hierarchical layers of Incident Command Systems (ICS) are engaged; specifically, a central entity (e.g., the federal government) typically coordinates the incident response allocating resources to different sites, which then get distributed to the affected by local entities. 
The benefits of an allocation decision to the different sites are generally not expressed explicitly as a closed-form utility function because of the complexity of the response and the random nature of the underlying phenomenon we try to contain.
 We use the classic approach of decomposing the NUM formulation and applying a primal-dual algorithm to achieve optimal higher-level decisions under coupled constraints while modeling the optimized response to the local dynamics with deep reinforcement learning algorithms. 
    The decomposition we propose has several benefits: 
1) the entities respond to their local utilities based on a congestion signal conveyed by the ICS upper layers; 
2) the complexity of capturing the utility of local responses and their diversity is addressed effectively without sharing local parameters and priorities with the ICS layers above; 
3) utilities, known as explicit functions, are approximated as convex functions of the resources allocated;
4) decisions rely on up-to-date data from the ground along with future forecasts.

\end{abstract}

\section{Introduction}
This paper's main contribution is a novel Network Utility Maximization (NUM) framework to address complex and heterogeneous resource allocation problems for hierarchical systems, where the objective is maximizing social utility. The application we study is Incident Command Systems (ICS).
During an incident, resources and services from first responders are necessary to contain its impact. These resources and services are made available by various entities with limited supply at any given time. 
ICS are typically hierarchical, whereby higher layers are engaged only if the lower layers cannot manage the response alone. As the scale and frequency of health crises and natural disasters continue to rise, it is increasingly common to require a higher layer of ICS to coordinate resource allocation.  
In this context, optimal (or near-optimal), foresighted and practical decision-making is critical to study beforehand. 

Utilizing a distributed primal-dual formulation proposed solution is decomposed so that it is possible to coordinate the management of wide-area crises. Primal-dual decompositions of NUM formulations are similar to a {\it free-market} mechanism: the higher layers communicate a congestion signal, akin to a shadow price for each resource, to the lower layers. The latter adds the resource cost in maximizing their local utility. In turn, the information about the local situation and priorities does not have to be explicitly shared, only the final demand for resources to reiterate the congestion signal. We can even use different approaches to address local complex dynamic optimal responses based on their specific geography, needs, or priorities.  

What distinguishes our formulation from the more conventional NUM formulations is that we consider cases where one needs to allocate resources facing a stochastic dynamic environment. Therefore, each resource allocation decision should account for not only the short-term but also the long-term utility. While we decompose the NUM problem in space and layers, the resource allocation problem is not decomposable in time, and the local sub-problems fall into the class of stochastic dynamic optimization problems. For instance, a relatively limited allocation of firefighter units to a wildfire may cause the fire to spread vastly, requiring substantially more resources to control it in the future. Therefore, we need foresighted decisions that use future forecasts built upon the present real-time data. A necessary condition for decomposability across localities and layers is that the allocations do not directly affect the utility of other sites except for the scarcity of the cumulative resources for each site contends. In reality, this is often an approximation but reflects the local nature of the response on the ground. 
Not just during crises but also for planning, one can adopt the proposed framework to shed light on the cost of potential disasters for a given availability of resources, even when all parties involved are responding optimally.

We study the proposed approach in two compelling scenarios: pandemic response and wildfire response. In these scenarios, we model a pandemic evolution or a wildfire propagation as Markov Decision Processes (MDPs), where we obtain the approximate optimal local response for an allocation via deep reinforcement learning algorithms. These algorithms rely on trial-and-error type training on agent-based simulation systems, where the necessary training is highly dependent on the scale of the simulations and possible action spaces.
Therefore, our divide-and-conquer strategy with the decomposition on the higher layer is significantly helpful in obtaining multiple local scalable reinforcement learning systems. In return, the higher layer does not require data/simulation model parameters or any other complexities and only establishes a market mechanism to find the best allocation to different localities. 

We run these agent-based simulations over a finite time horizon window starting from the present to obtain foresighted allocations. The initial state of MDP (or the simulation) is the most recent data from the ground. When the local ICS gathers more data, that new current time becomes the time zero, and we repeat the simulation and the overall optimization. Therefore, in a {\it shifting time window} manner, we find foresighted optimal allocations while using up-to-date ground data as simulation initializations. During that process, even though we make decisions for a future time window, only the near-decisions are finalized, while future ones get revised potentially based on new incoming data from the ground.    

\subsection{Related Work}

NUM framework establishes an analytical foundation to design distributed and modularized control of networks through investigating resource-constrained utility maximization problems. While initial works mostly use dual decomposition-based distributed algorithms, as in the seminal works~\cite{kelly1998rate, low1999optimization}, later various alternative decompositions of network resource allocation problems have been developed. We refer the interested reader to the tutorial paper~\cite{palomar2006tutorial} in this area. More recently, a multi-layered hierarchical NUM~\cite{karakocc2020multi, karakoc2018multi} and a fully-decentralized federated variant~\cite{karakocc2020federated} have been investigated, whereby in~\cite{karakocc2022federated} linear convergence rate for NUM algorithms (when the resource constraints are linear equalities) has been shown without the time-scale separation assumption between the iteration layers. An overwhelming majority of this literature considers a static setup, whereas in this paper we consider a dynamic problem where underlying systems evolve with the allocated resources in time. Among those considering a dynamic allocation, a common approach is to use Lyapunov drift-plus-penalty approach if the problem has soft intertemporal constraints~\cite{neely2003dynamic, wang2019multi, ferrari2018layered}. In the case of delivery contracts, where feasibility should be guaranteed with a deadline, ~\cite{trichakis2008dynamic} provides an exact solution with distributed implementation for the case where all the constraints in time are known exactly, and an approximate solution based on model predictive control where there is uncertainty in the future constraint set. To improve the convergence speed in this setup, a Newton method-based distributed algorithm has been proposed in~\cite{wei2012distributed}. For the case where the utility functions at localities change following a known set of deterministic dynamical equations,~\cite{ma2019optimal} proposes an iterative method based on dual decomposition. This setup is similar to our consideration. The differences lie in the fact that we consider large-scale disaster response problems with underlying dynamics following stochastic computer simulations. The decomposed problems at localities, in our consideration, are stochastic and complex and require approximate solutions such as deep reinforcement learning rather than being deterministic closed-form problems.

The paper is organized as follows. Next, we discuss the general modeling of a stochastic dynamic network utility maximization (NUM) problem for optimal resource allocation in an incident response followed by the proposed distributed solution approach. In the later sections, we present the details of the considered case studies. Finally, we end with numerical results. 

\section{Problem Formulation}
We formulate the resource allocation problem as maximizing the sum utility of independent MDPs subject to a common resources constraint. The notation is as follows:
\begin{enumerate}
    \item ${\cal P}^{\ell}$ is the population in location $\ell=1,\ldots, L$.
    \item $s^\ell_t$ is the state of $\ell$ at future horizon window time $t = 1,\ldots,T$. These states contain all the available information regarding how the disaster can impact locations, including information regarding the entities $p \in {\cal P}^{\ell}$ as well as the environmental factors.    
    \item $s^\ell_0$ is the state of $\ell$ at present (or $t=0$), i.e., it denotes the most recent available data from the ground. 
    \item ${\cal S}$ is the set of all possible state values, i.e., the state space. 
    \item $a_{p,t}^\ell$ is vector of resources allocated to $p$ at time $t$ in $\ell$. 
    \item ${\cal X}_{p,t}^\ell$ is the set constraining the individual resources (e.g., non-negativity).
    \item  $a^\ell_t$ is the collection of $a^\ell_{p,t}$, $\forall p$ for a given $\ell,t$.  It denotes an allocation decision, i.e., an allocation action in the MDP framework. 
    \item $y^\ell$ represents the resources assigned to location $\ell$ per unit of time, i.e., $\sum_{p\in{\cal P}^{\ell}} a_{p,t}^\ell \leq y^\ell$, $\forall t$.
    \item $A_{s^\ell_t}$ denotes the set of all available actions from state $s^\ell_t$, i.e., it is the action space from $s^\ell_t$.
    \item We assume that the disaster dynamics are Markovian, i.e., they can be defined with state transition probabilities $\mathbb{P}(s^\ell_{t+1} | s^\ell_t, a^\ell_t)$. We denote these dynamics with $W$, such that $s^\ell_{t+1} = W(s^\ell_t, a^\ell_t)$, mapping the previous state and the action pair to the new state probabilistically based on the underlying dynamics that define the evolution of the specific disaster. 
    \item It is common to have a probabilistic policy in stochastic dynamic optimization. $\pi^{\ell}(a^\ell_t| s^\ell_t)$ denotes a potentially probabilistic mapping from state space to action space, i.e., $\sum_{a^\ell_t \in A_{s^\ell_t}} \pi^{\ell}(a^\ell_t| s^\ell_t) = 1$, and $\pi^{\ell}(a^\ell_t| s^\ell_t) \geq 0, \forall \ell, t$.  
    \item ${\cal U}^\ell_p(a_{p,t}^\ell | s^\ell_t)$ is the immediate utility for $p$, with the allocation $a_{p,t}^\ell$ at state $s^{\ell}_t$.
    \item $z$ represents the total resource supply per time unit. This leads to the constraint $\sum_{\ell=1}^L y^\ell \leq z$, which couples the resources across all locations.
\end{enumerate}

The NUM problem 
formulation captures the basic problem of assigning resources under congestion/capacity constraints to maximize the sum utility for the population, i.e.:
\begin{align}
    \max_{\{y^\ell\}, \{\pi^\ell\}} &\sum_{\ell=1}^L  ~\mathbb{E}_{\pi^\ell}\left[ \sum_{t=0}^{T} \sum_{p\in{\cal P}^{\ell}} \gamma^t {\cal U}^\ell_p(a_{p,t}^\ell | s^\ell_t) \right]\label{eq:sum-utility}
   \\
   \mbox{s.t.}~~ & a^\ell_t \in A_{s^\ell_t},  
   ~~\forall \ell,t 
   \\ 
   &\pi^{\ell}(a^\ell_t| s^\ell_t) \geq 0,~~ \sum_{a^\ell_t \in A_{s^\ell_t}} \pi^{\ell}(a^\ell_t| s^\ell_t) = 1, ~~\forall \ell, t \\ 
   &a_{p,t}^\ell\in {\cal X}_{p,t}^\ell,~~ \sum_{p\in{\cal P}^{\ell}} a_{p,t}^\ell\leq y^\ell, ~~\forall \ell,t ,\label{eq:demand-supply-site}\\
    &\sum_{\ell=1}^L y^\ell\leq  z, ~~ \label{eq:demand-supply-total}\\
    &s^\ell_{t+1} = W(s^\ell_t, a^\ell_t),~~\forall \ell,t \label{eq:dynamics}
\end{align}
where $\gamma$ is a utility (or reward) discount factor, capturing the penalty for delaying the allocation of resources (commonly, $0 < \gamma \leq 1$). The expectation (denoted by $\mathbb{E}[\cdot]$) is over the sources of the stochasticity in the system as well as the randomized policy $\pi^\ell, \forall \ell$.
To ensure that the demand met is feasible, i.e., less or equal to the supply, we have constraint sets \eqref{eq:demand-supply-site}, \eqref{eq:demand-supply-total}. Finally, \eqref{eq:dynamics} accounts for the impact of the incident dynamics.  

These dynamics start with present-time data $s^\ell_0$, and we simulate the future states $s^\ell_t$'s via agent-based simulations. These simulations, which produce probabilistic future realizations branched off from $s^\ell_0$ by moving agents in the system, establish a playground for deep reinforcement learning algorithms to train and reach effective foresighted allocation strategies. As time moves forward and new data becomes available (at time $\tau$ for example), it replaces $s^\ell_0$ and the horizon window shifts, meaning $t=\tau$ becomes $t=0$, and so on. The simulations are then restarted from the updated $s^\ell_0$ and decisions are revised accordingly. Consequently, the higher-level allocations $y^\ell$ for all $\ell$ are updated only at a slower pace, such as at time intervals of $\tau$. For simplicity, in a single simulation run, it is assumed that $y^\ell$ remains constant for the duration of the future horizon window.   

There might be a need for more than one type of resource. In that case, we can assume that $a_{p,t}^\ell, y^\ell$ and $z$ are vectors where each entry is the quantity of a particular item type to be allocated or supplied.   
The division of the population over $L$ sites affected by the crisis here is known, and we assume that the dynamics of the event are independent across localities, i.e., they are disjointed MDPs. 

We assume that each of the $L$ sites has its own local ICS that is activated to address the local resource allocation problem, i.e., choose the values of $a_{p,t}^\ell$ for given $y^\ell$. 
Owing to the hierarchical nature of ICS, our tenet is that we can cast the optimal resource allocation problem for various disaster response scenarios in general in such a formulation, and the discriminating aspect lies in the resource type and, more importantly, in the dynamics that depend on the type of incident. 
The set of constraints in~\eqref{eq:dynamics} refers to the underlying dynamics of the incident in consideration. For instance, if the incident is a natural disaster, such as an earthquake, the destruction peaks at time zero and is generally non-increasing. In a pandemic, instead, the additional exposure of people causes waves of infections. 

Now, this view of the problem reflects the task of allocating resources to local ICSs when the event footprint is so vast that entities and suppliers at a higher layer need to be activated. However, it does not capture the information bottlenecks and the coarse information available about the individuals' utilities.
Even though it might be sometimes possible to estimate the impact of an incident over populations with a well-defined set of equations, in other cases, the way to tackle the response locally and the best practices vary significantly. By decomposing the problem into localities and by creating a market-like mechanism for the allocation of global resources to the local ICSs, where global ICS only uses the result of local optimization problems regardless of their complexity or choice of solution, our approach allows significant flexibility in terms of compatibility with different applications with different modeling of underlying dynamics as well as different solution methods. 

The real-life optimality of the response depends on various factors, first and foremost on how well localities gather data to gain situation awareness about what is unfolding and, second, how they use the information to decide what to do with the resources optimally. In this paper's case studies, we consider two dynamic incident models for $s^\ell_{t+1} = W(s^\ell_t, a^\ell_t),~~\forall t$ and assume information regarding the situation is perfect (the state $s^\ell_0$ is available, the state $s^\ell_t$ is observable through simulations, and the dynamics are known); our focus is on exemplifying how to obtain a strategy to use the resources optimally.
Another imprecision in our model is that it does not capture the external impact a local response may have in increasing the utility for other sites. For example, wildfires and epidemics may not be confined to localities, and the allocation in one place can reduce the risk for others since both can spread from one site to the other. However, it is mostly the case that for local ICSs, the other localities are more of an afterthought.

\section{Distributed Solution via Primal Dual
\\ Decomposition Theory}
\subsection{Decomposing the Global Problem}
\begin{figure}[t]
\centering
\resizebox{\linewidth}{!}{
\begin{tikzpicture}[scale=0.6]

\draw[fill,lightgray](2,2) rectangle (4,4);
\draw (2,2) rectangle (4,4);
\draw[fill,lightgray](1,-2) rectangle (3,0);
\draw (1,-2) rectangle (3,0);

\draw[fill,lightgray](-4,-2) rectangle (-2,0);
\draw (-4,-2) rectangle (-2,0);
\draw[fill,lightgray](6,-2) rectangle (8,0);
\draw (6,-2) rectangle (8,0);

\draw[line width=0.4mm,->] (3,1.8) -- (-2.8,0.2);
\draw[line width=0.4mm,->] (3,1.8) -- (2,0.2);
\draw[line width=0.4mm,->] (3,1.8) -- (7,0.2);

\draw[font=\bf]  (4.5 cm,-1 cm) node[anchor=center] {$\ldots$};
\draw[font=\bf]  (-0.5 cm,-1 cm) node[anchor=center] {$\ldots$};

\draw[font=\bf]  (3 cm,3 cm) node[anchor=center] {Center};
\draw[font=\bf]  (-3 cm,-1 cm) node[anchor=center] {$\ell = 1$};
\draw[font=\bf]  (2 cm,-1 cm) node[anchor=center] {$\ell$};
\draw[font=\bf]  (7 cm,-1 cm) node[anchor=center] {$\ell = L$};

\draw[font=\bf]  (-10 cm,-1 cm) node[anchor=center] {Local Optimizations: };

\draw[font=\bf]  (-10 cm,3 cm) node[anchor=center] {Main Problem: };

\draw[font=\bf,blue]  (-1.3 cm,1.2 cm) node[anchor=center] {$\lambda$};
\draw[font=\bf,blue]  (1.9 cm,0.9 cm) node[anchor=center] {$\lambda$};
\draw[font=\bf,blue]  (6.1 cm,1.1 cm) node[anchor=center] {$\lambda$};

\draw[line width=0.4mm,->,red] (-2.4,0.2) -- (2.8,1.6);
\draw[line width=0.4mm,->,red] (2.2,0.2) -- (3.1,1.6);
\draw[line width=0.4mm,->,red] (6.7,0.2) -- (3.1,1.6);

\draw[font=\bf,red]  (0.8 cm,0.6 cm) node[anchor=center] {$y^{1*}$};
\draw[font=\bf,red]  (3.2 cm,0.6 cm) node[anchor=center] {$y^{\ell*}$};
\draw[font=\bf,red]  (4.7 cm,0.6 cm) node[anchor=center] {$y^{L*}$};

\end{tikzpicture}}
\vspace{-2mm}
\caption{Message passing for the higher-layer allocations}
\label{figDecomp}
\end{figure}
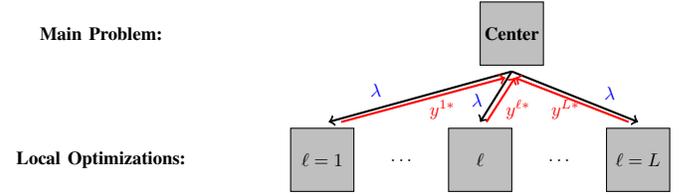

In this section, we apply decomposition theory, and in particular, the primal-dual decomposition, to the problem in \eqref{eq:sum-utility} and obtain a distributed solution that reduces the burden of exchanging data from the local sites to the central one. That is possible because the Markov Decision Process (MDP) dynamics in each location are modeled independently from each other and coupled only due to the total resources, $z$, being finite. 
That is, for a given $y^\ell$, the $a_{p,t}^\ell,~p\in {\cal P}^{\ell}$ values affect only the utilities inside location $\ell$. It follows that for a given  $y^{\ell}$, each location $\ell$ can solve the inner problem:
\begin{align}
      F_\ell (y^{\ell}) &= 
      \max_{ \pi^\ell} \mathbb{E}_{\pi^\ell}\left[ \sum_{t=0}^{T} \sum_{p\in{\cal P}^{\ell}} \gamma^t {\cal U}^\ell_p(a_{p,t}^\ell | s^\ell_t) \right]\label{eq:inner-problem}
   \\
   \mbox{s.t.}~~ & 
   a^\ell_t \in A_{s^\ell_t}, ~~\forall t \nonumber
   \\ 
   &\pi^{\ell}(a^\ell_t| s^\ell_t) \geq 0, \sum_{a^\ell_t \in A_{s^\ell_t}} \pi^{\ell}(a^\ell_t| s^\ell_t) = 1, ~~\forall t \nonumber\\ 
   & a_{p,t}^\ell\in {\cal X}^{
   \ell}_{p,t},~~\sum_{p\in{\cal P}^{\ell}} a_{p,t}^\ell\leq y^\ell, ~~\forall t ,\nonumber\\
    &s^\ell_{t+1} = W(s^\ell_t, a^\ell_t),~~\forall t, \nonumber
   \end{align}
where $F_\ell (y^{\ell})$ denotes the resulting optimal utility attainable for a given allocation  $y^{\ell} $ starting from present-time state $s^\ell_0$. Under the congestion constraint, the problem to solve for the higher layer: 
\begin{align}\label{eq:NWProb}
    \max_{y^{\ell}, \forall \ell}&~~\sum_{\ell=1}^L F_\ell (y^{\ell})~~~
    \mbox{s.t.}~~~ \sum_{\ell=1}^L y^\ell\leq  z.
\end{align}
This formulation corresponds to a NUM resource allocation~\cite{chiang2007layering}, which can leverage the dual decomposition.
Let $\lambda$ denotes the Lagrange multiplier. The Lagrangian of the upper layer of ICS is as follows 
\begin{equation}
    \mathcal{L} (\{y^{\ell}\},\lambda) = \sum_{\ell=1}^L F_\ell (y^{\ell}) - \lambda^\top \bigg[\sum_{\ell=1}^L y^\ell- z\bigg],
\end{equation}
and the dual problem to solve becomes
\begin{equation}
    \min_{\lambda} \max_{y^\ell, \forall \ell} \sum_{\ell=1}^L F_\ell (y^{\ell}) - \lambda^\top \bigg[\sum_{\ell=1}^L y^\ell- z\bigg].
\end{equation}
For fixed $\lambda$, the primal variables $y^\ell$ can be found as
\begin{equation} \label{yIter}
    y^{\ell*}(\lambda) = \arg \max_{y^{\ell}} \bigg\{ F_\ell (y^{\ell}) - \lambda^\top y^\ell\bigg\}\end{equation}
and we find the optimum dual variables via an iterative gradient descent method which the following update:
\begin{equation} \label{lambdaIter}
    \lambda^{k+1} = \bigg[\lambda^k + \alpha \bigg(\sum_{\ell=1}^L y^{\ell*}(\lambda^k)- z\bigg)\bigg]^+, k=1,2,\ldots
\end{equation}
where $\alpha$ denotes a step size, $k$ denotes iteration index and $[\cdot]^+$ denotes projection onto non-negative orthant and $y^{\ell*}$ is the solution of problem \eqref{yIter} evaluated based on the current iteration of $\lambda^k$. As illustrated in Fig.~\ref{figDecomp}, to sum up, the distributed solution finds $y^{\ell*}$ in each location $\ell$ separately for a given $\lambda$ whose value is then updated using new $y^{\ell*}$ from all locations. This process continues iteratively till a convergence criterion (market equilibrium) is satisfied. The $\lambda$ values serve as pseudo-prices, which are determined based on the total supply $z$ and the total demand $\sum_{\ell=1}^L y^{\ell*}$. 

In this method defined with Eqns.~\eqref{yIter}-\eqref{lambdaIter}, the convergence to a global optimum is guaranteed if $F_{\ell}(y^{\ell})$ is an increasing and differentiable concave function~\cite{karakocc2020multi}. However, in our case, rather than being a well-defined closed-form function, $F_{\ell}(y^{\ell})$ is the result of a challenging optimization problem, the local stochastic optimization in \eqref{eq:inner-problem}. Note that, here, our algorithm that solves the global allocation problem cooperatively, with~\eqref{lambdaIter}, does not use any information from the local optimizations except the returned total utility $F_{\ell}(y^{\ell})$ for given $y^\ell$. Therefore, as a result, it is unaffected by the choice of optimization methods in every locality, including if one chooses to use a deterministic policy or a randomized policy to solve the dynamic sub-problem or any heuristic approximation as long as they achieve near-optimal results.  

We highlight that $F_{\ell}(y^{\ell})$ is affected heavily by the initial state $s^\ell_0$, i.e., real-time data. For example, in the case of wildfire, our simulations do not create out-of-nowhere new fires, only model the propagation of existing ones. As a result, if there is no ongoing wildfire in a location, the utility of any firefighting resources will be zero. The utility, in this case, is a function of many factors in the real-time data, such as wind speed and direction, exact fire location, et cetera, that affect the simulation branches. Therefore, we need to revise $F_{\ell}(y^{\ell})$ values as well when we shift the time horizon window with incoming data in time.

\subsection{Solving the Sub-Problems in Localities} 
Owing to the complexity of solving the problems as in~\eqref{eq:inner-problem}, many different approximation methods have been proposed in the literature to reach sub-optimal allocations. 
The modular nature of our decomposition provides flexibility in choosing how to approach the solution of the inner local optimization. In fact, the gradient step at the higher layer only requires the value of the optimum demand in the local problem $y^{\ell*}$ in each location $\ell$ allowing different localities to use different methods to determine $y^{\ell*}$. The decomposition separates the solution process of locations except for the exchange of $\lambda$, which is analogous to a shadow price for the total resource allocation. 

The inner problem in~\eqref{eq:inner-problem} is a stochastic dynamic optimization modeled as an MDP. To solve such problems, traditionally dynamic programming and reinforcement learning, and recently, more commonly, deep reinforcement learning (which utilizes neural networks) approaches are being used due to the proliferation of computational advancements and new techniques. In this paper's case studies, we choose to use deep reinforcement learning algorithms along with agent-based simulations to solve the local problems in~\eqref{eq:inner-problem} due to their capacity to handle relatively large-scale complex problems. Nonetheless, our framework is flexible to accommodate the preferred method among dynamic optimization/reinforcement learning algorithms for the application in mind. We refer interested readers to~\cite{bertsekas2019reinforcement, sutton2018reinforcement} for detailed background on different algorithms to solve dynamic optimization problems.    

\subsection{Concave and Non-decreasing Interpolation}\label{interpolate}
The critical aspect regarding the solution of these local complex dynamic problems is that finding $F_{\ell}(y^{\ell})$ for a given $y^\ell$ is non-trivial. If a time-consuming computational model is available to find a $F_{\ell}(y^{\ell})$ for a given upper-layer allocation $y^\ell$ (i.e., to solve the inner dynamic problem), an offline computation might be preferable to an online one. In such cases, we may want to estimate $F_{\ell}(y^\ell)$ from samples $F_{\ell}(\mathrm{y}_i)$ obtained from numerical or experimental evidence of what is attainable with a vector $\mathrm{y}_i$ of possible values of $y^\ell$, and interpolate the response to the set of all possible values.  

The option we propose is to resort to a piece-wise linear interpolation. Let us denote the curve we fit to the samples with $\hat{F}_\ell (\mathrm{y})$, where $\mathrm{y}$ denotes a realization of $y^\ell$. Assume one has $n$ samples of $F_\ell (\mathrm{y}_i)$ for certain allocations $\mathrm{y}_i$, $i=1,\ldots,n$. We can cast the interpolation into an optimization that minimizes the mean square error from the sample points under the constraint that $\hat{F}_\ell (\mathrm{y})$ is non-decreasing and concave, which is due to the observation that additional resources do not reduce the utility and generally provide diminishing returns. Thus, we can formulate the  constrained curve fitting problem:
\begin{equation}\label{eq:interpolation}
    \begin{split}
        \min& \sum_{i=1}^n \left( \hat{F}_\ell (\mathrm{y}_i) - F_\ell (\mathrm{y}_i) \right)^2\\
        \text{s.t. }&  \hat{F}_\ell: \text{concave and non-decreasing}
    \end{split}
\end{equation}

If $\hat{F}_\ell$ is concave and non-decreasing, then
\begin{align}
    \hat{F}_\ell(\mathrm{y}) &\leq \hat{F}_\ell(\mathrm{y}_0) + \nabla \hat{F}_\ell^\top(\mathrm{y}_0) (\mathrm{y} - \mathrm{y}_0), ~\forall \mathrm{y},\mathrm{y}_0
\\
    \nabla \hat{F}_\ell(\mathrm{y}) &\geq 0, ~\forall \mathrm{y},
\end{align}
where $\nabla \hat{F}_\ell(\mathrm{y})$ is the gradient.

To simplify the notation, let us omit the location index $\ell$ and define:
\begin{equation}
    \mathbf{g}_i \triangleq \nabla \hat{F}_\ell(\mathrm{y}_i),~~~u_i \triangleq F_\ell (\mathrm{y}_i),~~~\hat{u}_i \triangleq \hat{F}_\ell (\mathrm{y}_i).
\end{equation}
We can write a problem to find the piece-wise linear approximation as follows: 
\begin{equation}
\begin{split}
        \min_{\hat{u}_i, \mathbf{g}_i, \forall i}& \sum_{i=1}^n \left( \hat{u}_i - u_i \right)^2 \\
        \text{s.t. }& \hat{u}_j \leq \hat{u}_i + \mathbf{g}_i^\top (\mathrm{y}_j - \mathrm{y}_i), ~\forall i,j\\
        & \mathbf{g}_i \geq 0, ~\forall i 
\end{split}
\end{equation}
which is a quadratic program (QP) since all the constraints are linear and the objective function is quadratic and, therefore, can be solved by one of the many available QP solvers.  
The optimization outcome is $n$ lines that pass through $\hat{u}_i$'s with slope $\mathbf{g}_i$ with minimum distance to the data points. Then, we can approximate the utility function $F_\ell (\mathrm{y})$ at point $\mathrm{y}$ with: 
\begin{equation}
    \hat{F}_\ell(\mathrm{y}) = \min_{i=1,..,n} \left\{ \hat{F}_\ell(\mathrm{y}_i)  + \nabla \hat{F}^\top_\ell( \mathrm{y}_i) (\mathrm{y} - \mathrm{y}_i) \right\}.
\end{equation}

With this function, we can estimate the result of the local optimization for any possible $y^\ell$ on the run. It can be suitable for cases where solving the inner optimization online is impossible with the computation power available.   
\subsubsection{Optimality gap}
We want to characterize the difference between the optimal solution of the problem in consideration and the solution we reach using the approximate objective function $\hat{F}_{\ell}(\mathrm{y})$. In the latter, we reach the optimal solution using the dynamics of the primal-dual iterations \eqref{yIter} and \eqref{lambdaIter} with the approximate function $\hat{F}_\ell (y^{\ell})$, i.e.:
\begin{align}
    \hat{y}^{\ell*}(\hat{\lambda}) &= \arg \max_{y^{\ell}} \bigg\{ \hat{F}_\ell (y^{\ell}) - \hat{\lambda}^{\top} y^\ell \bigg\}\\
    \hat{\lambda}^{k+1} &= \bigg[\hat{\lambda}^k + \alpha \bigg(\sum_{\ell=1}^L \hat{y}^{\ell*}(\hat{\lambda}^k)- z\bigg)\bigg]^+, k=1,2,\ldots
\end{align}
to solve the problem defined in~\eqref{eq:NWProb}. Let's define $f(Y) = \sum_{\ell=1}^L F_\ell (y^{\ell})$ and $\hat{f}(Y) = \sum_{\ell=1}^L \hat{F}_\ell (y^{\ell})$, where $Y$ is a long vector keeping all $y^\ell$ values $\forall \ell$. Furthermore, let's denote the optimal solution with and without the function approximation as $\hat{Y}^*$ and $Y^*$, respectively. 

Let us refer to the maximum total approximation error over locations sum-utility at any value of vector $Y$ as $\varepsilon$, i.e., 
\begin{align} \label{eq:appErr}
    |\hat{f} (Y) - f (Y)| \leq \varepsilon, \quad \forall Y.
\end{align}

\begin{theorem}
Assume the sum-utility function $f(Y)$ is strongly concave, i.e., $- \nabla^2 f(Y) \succeq m_f I, ~\forall \ell$, with a positive $m_f$, and an increasing function. Then, we have 
\begin{equation}
    \norm{Y^* - \hat{Y}^*} \leq 2 \sqrt{\frac{\varepsilon}{m_f}}.
\end{equation}
\end{theorem}

\begin{proof}
The first important observation about the solutions $\hat{Y}^*$ and $Y^*$ of the resource allocation problem in~\eqref{eq:NWProb} is that they satisfy the constraints with equality since objective utilities are increasing with additional resources. Therefore, they are both feasible points, i.e., with or without the approximations.  

Notice that due to the definition of maximum, we can write 
\begin{equation} \label{eq:MaxS}
    f(Y^*) \geq f(\hat{Y}^*), \quad \hat{f}(\hat{Y}^*) \geq \hat{f}(Y^*),
\end{equation}
since the algorithms would converge to the other (feasible) points otherwise.

Using~\eqref{eq:appErr} and~\eqref{eq:MaxS}, we have 
\begin{equation}
    \begin{split}
      \hat{f}(Y^*) + \varepsilon \geq f(Y^*) \geq f(\hat{Y}^*) \geq \hat{f}(\hat{Y}^*) - \varepsilon,
    \end{split}
\end{equation}
and, therefore,  
\begin{equation}
    f(Y^*) - f(\hat{Y}^*) \leq 2\varepsilon. 
\end{equation}

Due to the strong concavity of $f(Y)$, 
\begin{equation}
    - f(Y^*) - \nabla f(Y^*)^\top(\hat{Y}^{*} - Y^*) +\frac{m_f}{2} \norm{ \hat{Y}^{*} - Y^*}^2 \leq - f(\hat{Y}^{*}).
\end{equation}

All of our constraints are linear, and sum-capacity constraints are over disjoint sets of variables (indices of $Y$), and therefore, allowable movement at the intersection of the set of linear equalities occurs at a hyperplane, where both $Y^*$ and $\hat{Y}^*$ lie on. Furthermore, at constrained maxima $Y^*$, $\nabla f(Y^*)$ should be perpendicular to this hyperplane since otherwise, we could increase $f(Y)$ by moving towards the direction of the projection of the gradient onto the hyperplane. Therefore, $\nabla f(Y^*)^\top(\hat{Y}^{*} - Y^*) = 0$. 
In such a case,
\begin{equation}
\begin{split}
         \frac{m_f}{2} \norm{ \hat{Y}^{*} - Y^*}^2 \leq f(Y^*) - f(\hat{Y}^*) \leq 2\varepsilon
\end{split}
\end{equation}

An exception to the orthogonality of the gradient at solution and hyperplane might occur if there are corner points due to additional constraints, such as the non-negativity of individual variables. Even if these constraints pull the solutions to the corner points, this mapping is non-expansive, i.e., $\norm{ \hat{Y}^{*} - Y^*}$ gets smaller or stays the same. Therefore, in general, we can conclude: 
\begin{equation}
    \norm{Y^* - \hat{Y}^*} \leq 2 \sqrt{\frac{\varepsilon}{m_f}}.
\end{equation}
\end{proof}
\section{Case Studies}

We consider two relevant disaster response scenarios today: wildfire and pandemic response. More specifically, we focus on the firefighting unit allocation in wildfires and vaccine allocation in pandemics. Even though there are many differences between these two scenarios in resource types, actions, and underlying dynamics, we can write both as our general formulation in Sec.~II. 

Before getting into the differences and the application-specific details, we first highlight the fact that the stochastic models for the dynamics $s^\ell_{t+1} = W(s^\ell_t, a^\ell_t),~~\forall t$ can in both cases be described as instances of stochastic multi-agent interactions over a particular graph, which in one case (the fire dynamics) represent the topological characteristics of the terrain and in the other (the pandemic) the social interactions among individuals and contagion.

\subsection{Propagation Through Multi-Agent Networks}\label{sec:propagation}

Assume we have a graph ${\cal G}^{\ell} = ({\cal P}^{\ell}, {\cal E}^{\ell})$, where each node $p \in {\cal P}^{\ell}$ denotes a person in the dynamics of a pandemic or a small forest area in a wildfire. In both scenarios, the edges represent the possible contacts that spread the phenomenon among neighboring nodes: the social contact causing the spread of disease from an infected individual, and the ignition of a neighboring zone due to the proximity to others that are burning. Each node is in only one of several possible discrete states: for pandemics, states can be susceptible, infected, recovered, vaccinated, or dead, and for wildfire, states are vulnerable, on fire, extinguished, or burnt.

Therefore, omitting index $\ell$ for simplicity, node $p$'s state at time $t$, $s_{p,t}$, is a categorical random variable that can change due to the outcome of interaction and the local state. 
If we normalize the time between decision/action epochs to be $1$, we can divide each decision period into infinitesimal intervals of time $dt\ll 1$, during which there can be zero or at most one interaction between a node and its neighbors with probability $0\leq w_{p,k}\leq 1$ proportional to $dt$ (two or more would be $o(dt^2)$) that can result in a change of state.
Let us assume that $1/dt$ is an integer number. 
Let ${\cal N}_p$ be the neighbors of node $p$ and ${\cal S}_p\subseteq {\cal N}_p$ be the set of neighbors of node $p$ that are susceptible. 
Over each of the $1/dt$ sub-intervals between two decision epochs each node $p$ can meet a neighbor $k$ with probability $w_{p,k}$, and infect/ignite of a susceptible neighbor $k\in {\cal S}_p \subseteq {\cal N}_p$, or make no contact, with probability $w_{p,p}=1-\sum_{k\in{\cal N}_p} w_{p,k} $.
Let us assume that:
\begin{enumerate}
    \item $s_{p,t}=0$ corresponds to susceptible/vulnerable state
    \item $s_{p,t}=1$ is infected/on fire
    \item $s_{p,t}=2$ is dead/burnt
    \item $s_{p,t}=3$ is vaccinated/extinguished 
    \item $s_{p,t}=4$ is recovered (this state does not exist in the case of wildfire).
\end{enumerate}
Let also introduce the random indicator variable $e_{p,k}$ such that:
\begin{align}
    e_{p,k}\triangleq 
    \begin{cases}
    1 & \mbox{if $p,k$ come in contact}\\
    0 &\mbox{else}
    \end{cases},~~~k\in {\cal N}_p 
\end{align}
and $e_{p,p}=1$ if $e_{p,k}=0, \forall k \in {\cal N}_p$, so that $\mathbb{E}[e_{p,k}]=w_{p,k}$. Also, we can introduce the node action $a_{p,t}$ such that $a_{p,t}=1$ if $p$ is vaccinated or the cell fire is extinguished at time $t$, and $a_{p,t} = 0$ else. 
For the pandemic, we can write:
\begin{align}\label{pandModel}
    s_{p,t+(i+1)dt}
    =
    \begin{cases}
    1 &\mbox{if }s_{p,t+i dt}= 0 \\
    &\mbox{~and~}(e_{p,k}=1 \mbox{~and~}  s_{k,t+i dt}= 1)\\
    & \mbox{~for any~} k\in {\cal N}_p;\\
    2 &\mbox{if }s_{p,t+i dt}= 1 \mbox{~and $p$ dies};\\
    3 & \mbox{if $i=0$},~s_{p,t}= 0 
    \mbox{~and~} a_{p,t}=1;\\
    4 & \mbox{if } s_{p,t+i dt}= 1
    \mbox{~and~$p$ recovers};\\
    s_{p,t+i dt} &\mbox{otherwise,}
    \end{cases}
\end{align}
for $i=0,\ldots,1/dt-1$, where the death and recovery probabilities, say $d_p$ and $r_p$, are nodal characteristics and $1-d_p-r_p$ is the probability that the agent remains infected. Note that 2, 3, and 4 are absorbing states. 
For the wildfire, the states evolve similarly, with the main difference that state 4 does not exist, and the fire gets extinguished while $s_{p,t}=1$:
\begin{align}\label{fireModel}
    s_{p,t+(i+1)dt}
    =
    \begin{cases}
    1 &\mbox{if }s_{p,t+i dt}= 0 \\
    &\mbox{~and~}(e_{p,k}=1 \mbox{~and~}  s_{k,t+i dt}= 1)\\
    & \mbox{~for any~} k\in {\cal N}_p;\\
    2 &\mbox{if }s_{p,t+i dt}= 1 \mbox{~and $p$ is burnt};\\
    3 & \mbox{if $i=0$},~s_{p,t}= 1 
    \mbox{~and~} a_{p,t}=1;\\
    s_{p,t+i dt} &\mbox{otherwise.}
    \end{cases}
\end{align}

\subsection{Case Study I: Pandemic Response}

\subsubsection{Background}
The first case study for our framework is that of pandemic response, which has gained significant attention due to the COVID-19 pandemic. Much of the research in this field centers on finding the best ways to distribute vaccines, but the same principles and techniques can also be applied to other resources, such as hospital beds and medical personnel, as well as determining the most effective policies for stay-at-home orders and travel restrictions.

In general, two different approaches are employed to model the infection dynamics. In \textit{compartmental} model, the population is divided into compartments (e.g., susceptible, exposed, infected, recovered), where each individual in these compartments is modeled as identical with others in terms of contact probability, etc. In contrast, in the \textit{agent-based simulation} model, the infections are simulated via computationally expensive methods where the individuals are modeled more heterogeneously, and the studies that use these models focus on comparing a number of predetermined vaccination strategies mostly~\cite{duijzer2018literature}.
In the paper~\cite{dalgicc2017deriving}, the authors compare the outcomes of optimal vaccine distribution under both compartmental and agent-based simulation models. They found that policies developed using more realistic agent-based simulations were more effective in reducing the overall number of infections. However, compartmental models are still commonly used because they are more amenable to mathematical optimization. Prior research in this field has mainly focused on influenza pandemics.
In~\cite{medlock2009optimizing}, the authors study which age group should be prioritized in influenza vaccinations; the result is that the optimal strategy prioritizes schoolchildren and adults aged 30 to 39 years. The work in~\cite{chen2014planning} proposes an optimization formulation for vaccine distribution in developing countries with detailed supply-chain constraints. In~\cite{enayati2020optimal}, the authors study optimal influenza vaccine distribution by considering heterogeneous populations with multiple subgroups based on geographic location and age, as well as fairness as a criterion. In that formulation, the authors model the population dynamics by also incorporating the contact between different age groups, and the optimization aims to keep the reproduction number (i.e., expected secondary infection rate) under $1$ with the minimum amount of vaccine use. In~\cite{eames2012measured}, the authors estimate social contact matrices among different age groups. Also related are ~\cite{zaric2001resource}, which studies epidemic control over short time horizons and ~\cite{duijzer2016most}, which investigates an optimal vaccine allocation problem over multiple populations, to maximize the benefits of herd immunity. For the other studies in this area, we refer the interested reader to the survey in~\cite{duijzer2018literature} and references therein. 
In recent studies, the attention naturally shifted towards COVID-19. In~\cite{chen2020allocation}, the authors investigate optimal static and dynamic allocation policies for COVID-19 vaccinations over different age groups and show that dynamic policies provide a significant improvement over static policies. In~\cite{bertsimas2020optimizing}, the optimal allocation over different geographic locations and age groups is investigated using an epidemiological model called DELPHI. The study uses real data from the United States and specifies which states and age groups should be prioritized. Finally,  \cite{babus2020optimal}~provides an analysis of COVID-19 vaccine prioritization based on age-based mortality rates and occupation-based exposure risks. 

\subsubsection{Modeling Details}
We employ agent-based simulations due to their capability of highlighting the heterogeneity of populations. In our simulations, every agent $p$ is characterized by a state variable, $s_{p,t}$, that indicates whether they are susceptible, infected, etc., as well as a fixed age-group variable that indicates which age group they belong to: teen, adult, or elderly. The simulations iterate random social interactions over a graph network based on the rules outlined in~\eqref{pandModel}. The death probability, $d_p$, of the adult group is set to be $10$ times higher than that of the teen group and $1/10$ of the elderly group. Recovery times, which are obtained from a Poisson distribution with a certain mean time after infections, do not vary by age group. The model is implemented using the Python library Mesa~\cite{python-mesa-2020, mesablog}.

\begin{figure}[t]
    \centering
    \includegraphics[width = \linewidth]{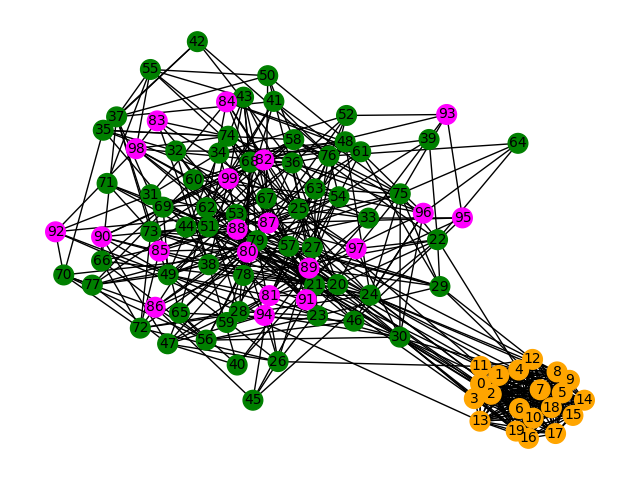}
    \caption{An example of a social graph for pandemic propagation, where orange nodes represent teenagers, green nodes represent adults, and purple nodes represent the elderly population. The numbers on the nodes denote their unique identifiers.}
    \label{fig:exampleGraph}
\end{figure}

The underlying social graphs are produced as follows. First, we form families between teens and some of the adult population, and connect all teens imagining a school setting. And later, we connect the elements in the union set of elderly nodes and adult nodes randomly in an Erdos-Renyi fashion. In addition, we ensure that each graph we use is connected, i.e., there are no isolated nodes. An example graph is illustrated in Fig.~\ref{fig:exampleGraph}.

Once the simulation tool is developed as described above, we can use it to solve the local problem~\eqref{eq:inner-problem}, the task of selecting which nodes to vaccinate with a total supply of $y^\ell$ per time unit. In this problem, $a_{p,t}^\ell$ denotes the action of vaccinating node $p$ at time $t$, and ${\cal X}^{\ell}_{p,t} = \{0,1\}$ for all $p,t,\ell$. The goal is to find $\pi^\ell(a^\ell_t| s^\ell_t)$, a randomized policy for determining which nodes to vaccinate by observing the state $s^\ell_t$, with the objective of maximizing total expected utility over a horizon length $T$. In this case study, we set the utility as negative of new deaths in location $\ell$ at $t$, i.e., the policy aims to minimize total (technically, discounted sum with $\gamma^t$ with $\gamma \approx 0.99$) death over $T$.   
  
To find a near-optimal policy to maximize total utility at $\ell$, we employ a deep reinforcement learning (DRL) model. To train our DRL model, we utilize Proximal Policy Optimization (PPO) algorithm with invalid action masking~\cite{huang2020closer}. PPO is a policy gradient method for reinforcement learning (see~\cite{schulman2017proximal} for details), and invalid action masking helps to ensure the model is taking feasible actions during training and execution. As the action space, we use the set of all nodes, where with the use of masking, the set of actionable nodes is reduced to only those that are in a susceptible state, i.e., we take infected, dead, recovered or already vaccinated nodes out. As for the neural network architecture, we use a single-layer Graph Convolutional Network (GCN)~\cite{kipf2016semi} and a single-linear layer as the shared feature extractor network, followed by 2-layer multi-layer perceptron (MLP) networks for both policy and value networks. As the activation function, ReLU is used throughout the model. The implementation is done in Python utilizing the libraries such as Stable Baselines 3~\cite{stable-baselines3}, OpenAI Gym~\cite{brockman2016openai}, and PyTorch Geometric~\cite{fey2019fast}.     

Recently, there have been papers~\cite{trad2022towards, wei2021deep, bushaj2022simulation, ling2024cooperating} that use DRL to study vaccine allocation. In these studies, the vaccination policy searched for by DRL is typically restricted to group-based allocation, such as age-group-based vaccination. While this is a practical strategy for limiting the action space and making DRL training easier, we believe that making each susceptible node a separate potential action utilizes the detailed heterogeneous simulation models more effectively. For example, with our model parameters where the elderly group has a 10x and 100x higher mortality rate than the adult and teenager groups, respectively, if we restrict the action space to selecting age groups and then vaccinating randomly within the group, the policy converges to an elderly-first vaccination policy since mortality rate directly impacts the total number of deaths significantly. However, according to our observations, it appears that our DRL model selects nodes based on their age group, their node degree (how connected they are to others), and their distance to infected nodes. The extra freedom given to the model in taking actions opens the door to more innovative strategies.

\subsubsection{System Integration and Numerical Examples} 
In this case study, we generate 5 locations with distinct underlying graphs and demographic characteristics as outlined in the Table~\ref{tab:locations}. The table shows the number of nodes in each group, as well as the average degree of the elderly group (EAD) and the average degree except for the teenager group (AD). The teenager group is excluded because of the high level of connectivity among them, which is a result of the school setting assumed during the graph creation.

\begin{table}[t]
    \centering
        \begin{tabular}{ |c|c|c|c|c|c| } 
        \hline
        Location & \# teen & \# adult & \# elderly & EAD & AD \\
        \hline
         Loc.1 & 20 & 50 & 30 & 7.56 & 8.24 \\ 
         Loc.2 & 30 & 60 & 10 & 6.7 & 7.97\\
         Loc.3 & 20 & 60 & 20 & 8.55 & 8.3\\
         Loc.4 & 20 & 60 & 20 & 8.7 & 9.05\\
         Loc.5 & 30 & 60 & 10 & 7.3 & 7.48\\
         \hline
        \end{tabular}
    \caption{Locations used in evaluation of pandemic response}
    \label{tab:locations}
\end{table}

\begin{figure}[t]
    \centering
    \includegraphics[width = \linewidth]{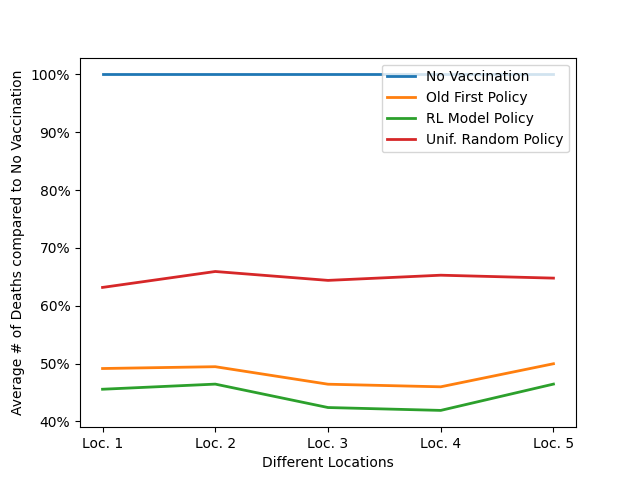}
    \caption{Performance of various policies at different locations.}
    \label{fig:PolvsLoc}
\end{figure}

In Fig.~\ref{fig:PolvsLoc}, we compare the performance of various policies, namely, no vaccination, old-first policy (which randomly vaccinates the elderly group first, adult group second, and finally teenager population), uniformly random selection among susceptible nodes, and our DRL model. All these policies are run for 50 time units, where each policy apply 1 vaccine per time unit. In addition, contact probability of two nodes sharing an edge is 0.02, death probability is $0.01$ each time unit for adult group, $0.1$ for elderly, $0.001$ for teen, and average recovery time is $14$ time units for all infected. We start all the experiments with 5 randomly infected nodes and run each policy 10000 times. On the y-axis, we plot the average number of deaths at the end of 50 time units, normalized with respect to the no vaccination policy. The results are consistent across different locations. Our results show that even if vaccinations are given out randomly, it still greatly lowers the number of deaths on average. The old-first policy, which is intuitively a good strategy (especially with our mortality rate parameters across different age groups), also provides a significant improvement compared to the random policy.
Our DRL model convincingly outperforms the old-first policy across all locations. It's worth noting that the RL models presented for each location are relatively simple, and each have been trained with random initialization (randomly infected nodes at the start) for about 90 minutes on a single Intel Xeon 6226R CPU.
When we train a model with fixed initialization (same nodes are infected at the start at each run) for the same amount of time, for example, training becomes easier and it achieves roughly 10 percent reduction in the average number of death (in the environment with the same fixed initialization) compared to the model trained with random initialization. There is room for improvement on these results with more training time, usage of parallel computing and GPUs, different network architectures etc. Since our aim here is to illustrate the proposed method in a case study, we select to use a DRL model fairly simple to reproduce and can handle different initializations.   

\begin{figure}[tb]
    \centering
    \includegraphics[width = \linewidth]{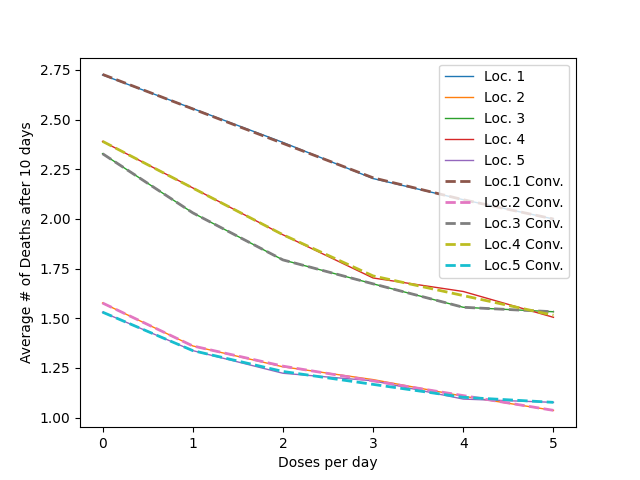}
    \caption{Negative utility vs allocated dose per day to a location along with convex piece-wise linear interpolations}
    \label{fig:deathvsdose}
\end{figure}

After training DRL models for each location, we can solve inner optimizations to obtain $F_{\ell}(y^{\ell})$ for given allocation $y^{\ell}$, i.e., we now know how to redistribute $y^{\ell}$ to the population at $\ell$ efficiently. $F_{\ell}(y^{\ell})$ in this case denote total utility achievable with $y^{\ell}$, which is the only value needed from local optimizations to solve the higher layer allocations. This value hides all the intricacies of complex dynamics at locality $\ell$. Next, we use our simulation tool and DRL model to generate samples of $F_{\ell}(y^{\ell})$ for different $y^{\ell}$ values. At this stage, we do not train our model any further, we only record its performance. Once we have collected these samples, we use concave and non-increasing interpolation to obtain a piecewise linear approximate utility function $\hat{F}_\ell (\mathrm{y})$ as described in Sec.~\ref{interpolate}. An example run is shown in Fig.~\ref{fig:deathvsdose}. In this example, we take samples for all possible values between 0 and 5, so we don't use the approximate function to interpolate but rather for convexification if some samples do not follow convex behavior.  

\begin{figure}[tb]
    \centering
    \includegraphics[width = \linewidth]{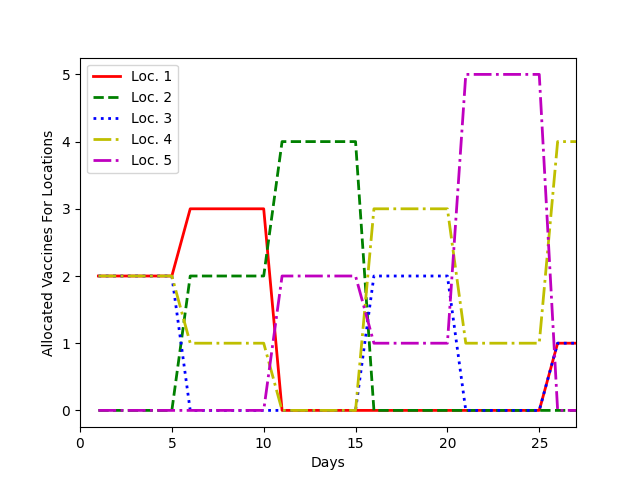}
    \caption{An example run of higher-layer allocations with $z = 6$}
    \label{fig:AlocvsIter}
\end{figure}

\begin{figure}[t]
\centering
\subfloat{
  \includegraphics[width=0.24\textwidth]{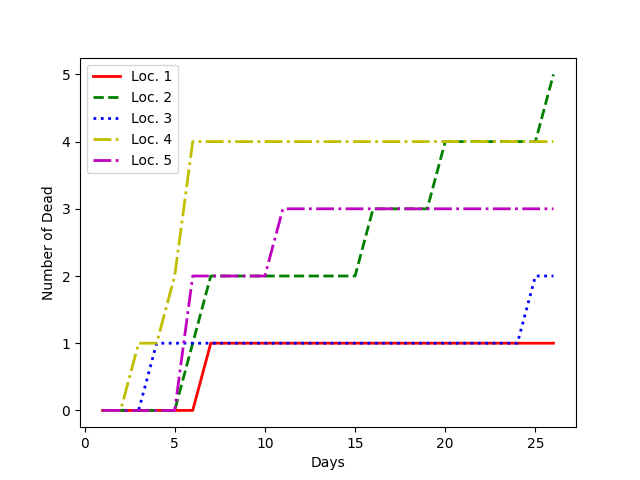}}
  \subfloat{
  \includegraphics[width=0.24\textwidth]{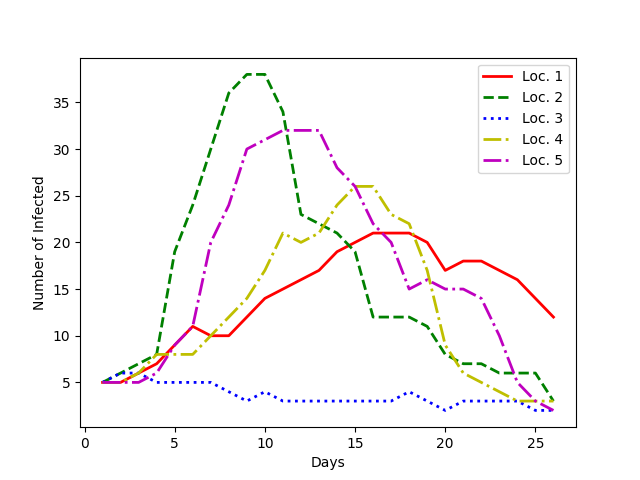}}
  \\
  \vspace{-4.2mm}
  \subfloat{
  \includegraphics[width=0.24\textwidth]{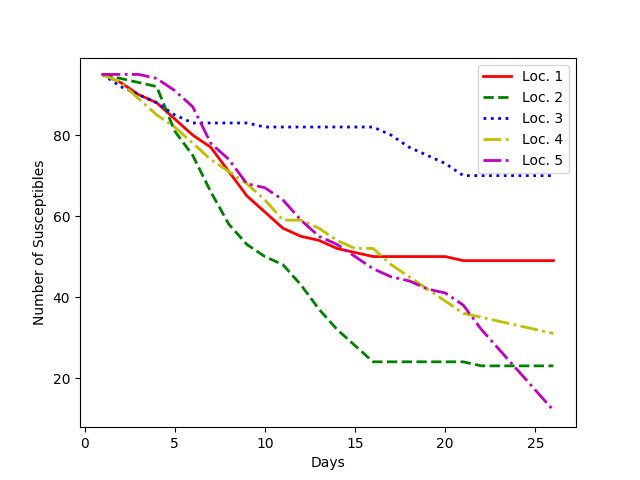}}
  \subfloat{
  \includegraphics[width=0.24\textwidth]{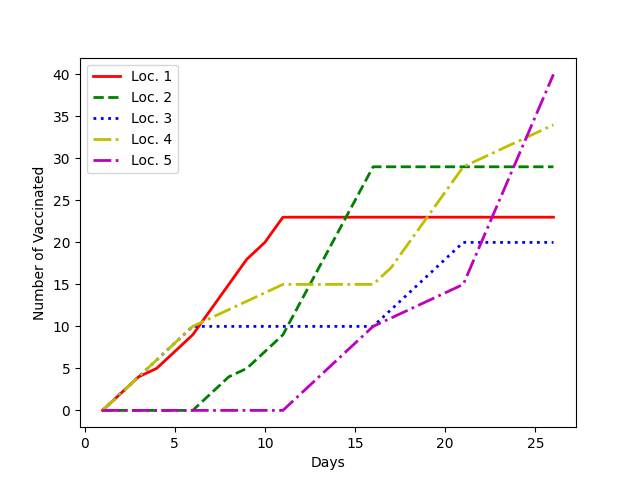}}
  \vspace{-1mm}
\caption{Accompanying stats to the run in Fig.~\ref{fig:AlocvsIter}}
\label{fig:AccStats}
\end{figure}

Finally, the results obtained from the local models are integrated utilizing the methodology outlined in Eqns.~\eqref{yIter}-\eqref{lambdaIter} in order to arrive at higher-layer allocations. An exemplary execution of the complete system, with a look-ahead horizon length of $10$ and higher-layer allocation update period of $5$ time units, is presented in Fig.~\ref{fig:AlocvsIter} and Fig.~\ref{fig:AccStats}. 
In a nutshell, the example system works as follows. In the beginning, randomly 5 nodes are infected at each location. By looking at a horizon of 10 days, the system makes higher-layer allocations, wherein resources are primarily directed towards locations with a higher concentration of elderly individuals. Subsequently, the local DRL systems employ these allocations to vaccinate nodes in accordance with their own models. After 5th day, local models observe their own location and make predictions based on this information for the next 10 days and share their $\hat{F}_\ell (\mathrm{y})$ after the convex interpolation. The upper-layer subsequently utilizes this information to calculate and respond with updated allocations, and this process is repeated iteratively on a rolling horizon basis.
When we run this setup repeatedly, we observe unique realizations and, as a result, unique higher-level allocations. Even though the allocations are skewed towards to locations with more elderly at the beginning as expected, the system adapts effectively to the underlying realizations and manages to keep death numbers under control. For instance, in the presented example, the system stops allocation to Loc. 1, which has highest elderly population, after tenth day completely, and redirects resources to the other locations upon observing that they are in a statistically more precarious situation.
It is important to note that, due to the utilization of numerous approximations, it is not possible to make claims of optimality with regard to the system under consideration. Nevertheless, the system presents an impression of efficient resource allocation according to our observations.

\subsection{Case Study II: Wildfire Response}

\subsubsection{Background}
As a second case study, we examine the problem of wildfire response. In this case, we aim to incorporate the dynamics of wildfire propagation across different regions as an input for optimizing response and resource allocation for extinguishing fires. To effectively address this problem, it is crucial to have a stochastic model for the evolution of wildfires in order to find the best policies through dynamic optimization.
Wildfire propagation has been widely studied in the literature. Different approaches consider various different factors affecting the fire spread, such as fuel type (type of vegetation), humidity, wind speed and direction, forest topography (slope and natural barriers), and fuel continuity (vegetation thickness)~\cite{alexandridis2008cellular}. The earlier works (e.g.,~\cite{rothermel1972mathematical}) primarily focused on developing dynamic equations that capture the physical nature of the propagation through controlled laboratory experiments.
When it comes to characterizing propagation over a large geographical area with more randomness, approaches can be classified into two types in terms of spatial representation: 1) models based on continuous planes, and 2) models based on grid representation. We adopt the latter in our modeling, due to its relative computational efficiency. A popular approach in grid-based methods is using a cellular automaton (CA) model. This method involves dividing a finite area into a large number of grid cells, where the evolution of the state in each cell is based on the state of its nearby neighbors and its local state through a set of rules that map these state values into transition probabilities. This discrete model allows for efficient computation and simulation while also accounting for the stochasticity of state transitions. Examples of works that use CA models include~\cite{karafyllidis1997model, alexandridis2008cellular, freire2019using}.

\subsubsection{Modeling Details}
Similar to the pandemic response, we again employ an agent-based simulation approach to evaluate wildfire response strategies. The simulation model employed is consistent with the methodology described in Section~\ref{sec:propagation}. The underlying graph structure of the simulation is a mesh grid neighborhood network, wherein each cell, denoted by $p$, possesses a state variable, denoted by $s_{p,t}$, that reflects its condition, which can be one of four possibilities: vulnerable, ignited, burnt, or extinguished. Additionally, each cell possesses unique characteristics, such as vegetation density and type, that are taken into account within the simulation. The agent-based simulation tool employed is able to simulate the dynamic evolution of these cell states, as described by the mathematical formulation presented in~\eqref{fireModel}. 

\begin{figure}[t]
    \centering
    \includegraphics[width = \linewidth, trim={0.8cm 0 3.6cm 0.5cm}, clip]{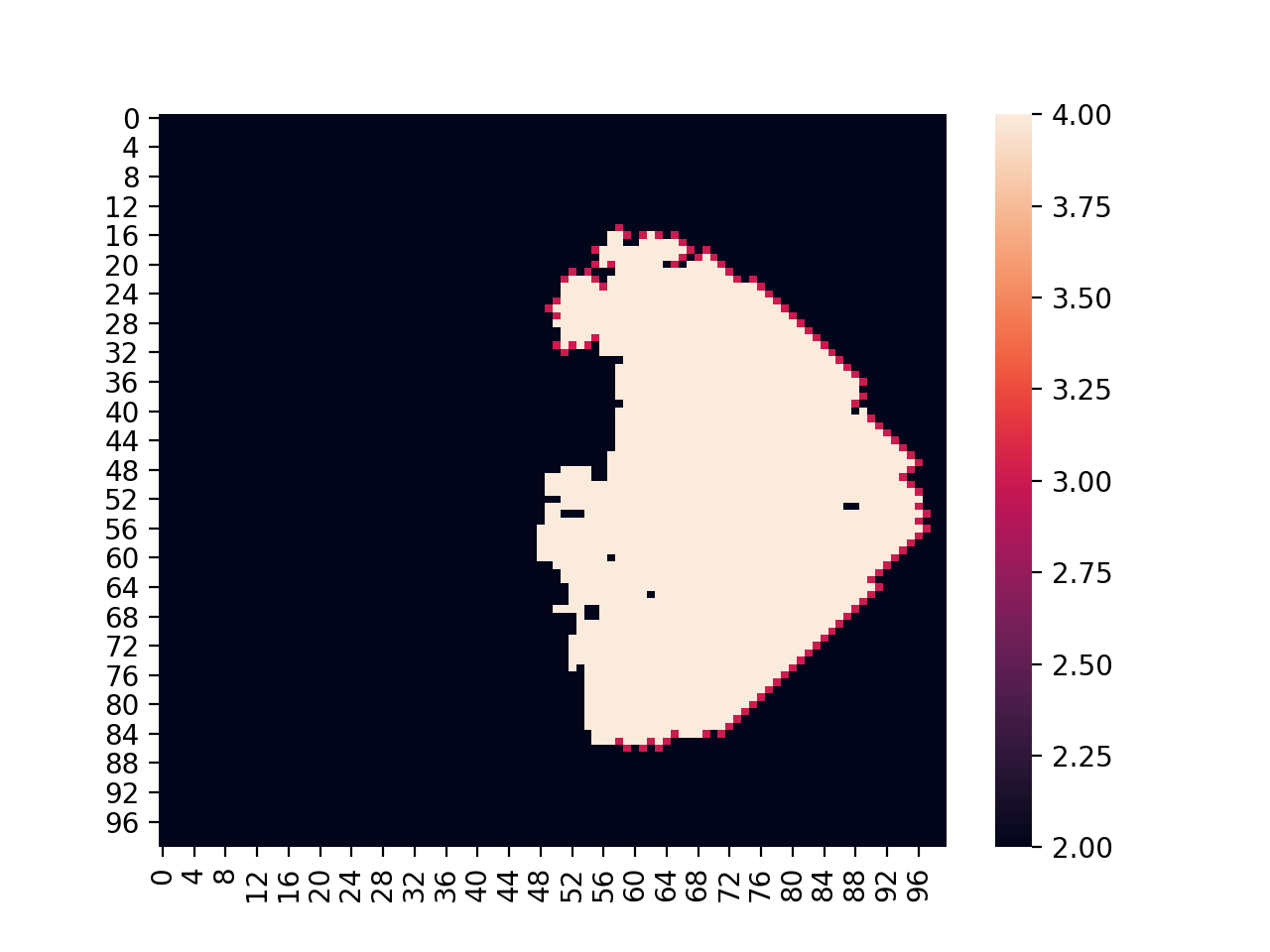}
    \caption{An example wildfire propagation, where light cells represent the burnt, dark cells represent the vulnerable, and red cells represent the ablaze. The size is $100\times100$.}
    \label{fig:exampleCA}
\end{figure}

The probability of cell $p$ being ignited due to the fire in neighboring cell $k$ (represented by $w_{p,k}$) is influenced by factors such as the vegetation type and density in cell $p$, wind speed and direction. We model this propagation probability using a multiplicative function that takes into account multiple factors, similar to the approach in~\cite{alexandridis2011wildland}:
\begin{equation}\label{eq.fire-spread}
\begin{split}
    w_{p,k} = \kappa (1+v_{p}) (1+d_p) \phi_{p,k}  
\end{split}
\end{equation}
where $\kappa$ represents a normalization constant, $v_p$ is a constant dependent on the type of vegetation present in cell $p$, $d_p$ is a constant dependent on the density of vegetation within cell $p$, and $\phi_{p,k}$ is a constant that is contingent upon the wind direction and speed. An illustration of a simulation run with strong eastward wind is shown in Fig.~\ref{fig:exampleCA}.

We use this simulation tool to address the inner problem outlined in~\eqref{eq:inner-problem}, which in the context of this case study pertains to the problem of determining the optimal cells to extinguish in order to maximize the discounted sum of utilities over a given horizon, given the allocation of firefighting units to location $\ell$, $y^\ell$.
In contrast to the pandemic response scenario, there are additional constraints on the action of extinguishing cell $p$, $a_{p,t}$, in the context of wildfire management. These constraints stem from the movement limitations of the firefighters. Specifically, we impose a restriction on the potential travel distance of a firefighting unit at each discrete time step, and the unit is only able to be at a specific location at each time. As a result, the problem is then can be formulated as identifying the optimal trajectories for the allocated firefighting units such that the cells that are on fire and situated within these trajectories are extinguished.

We incorporate the movement constraints directly into the simulation tool, ensuring that every action selected by the DRL model is a feasible movement. To determine the optimal movement of the firefighting units, we employ the PPO algorithm to train the DRL model. The neural network architecture utilized comprises of a two-layer Convolutional Neural Network (CNN) and a single linear layer serving as the shared feature extractor network. This is followed by 2-layer MLP networks for both the policy and value networks. The ReLU activation function is employed throughout the model. The implementation is carried out in Python using libraries such as Stable Baselines 3 and OpenAI Gym.

\subsubsection{System Integration and Numerical Examples} 
In this evaluation, we created two locations with different wind and vegetation characteristics, as shown in Table~\ref{tab:locationsWildfire}. The table displays the wind direction, speed (as a percentage of the typical maximum), and the normalized average vegetation coefficient (NAVegC). The NAVegC indicates the average impact of vegetation type and density on the rate of wildfire spread in a location, using a normalized scale. It's important to note that the vegetation density and type vary between cells, and the NAVegC is an average representation.

\begin{table}[t]
    \centering
        \begin{tabular}{ |c|c|c|c| } 
        \hline
        Location & Wind Dir. & Wind Speed & NAVegC \\
        \hline
         Loc.1 & $\leftarrow$ & $30\%$ & 1  \\ 
         Loc.2 & - & $0\%$ & 0.75 \\
         \hline
        \end{tabular}
    \caption{Locations used in evaluation of wildfire response}
    \label{tab:locationsWildfire}
\end{table}

\begin{figure}[tbh]
    \centering
    \includegraphics[width = \linewidth]{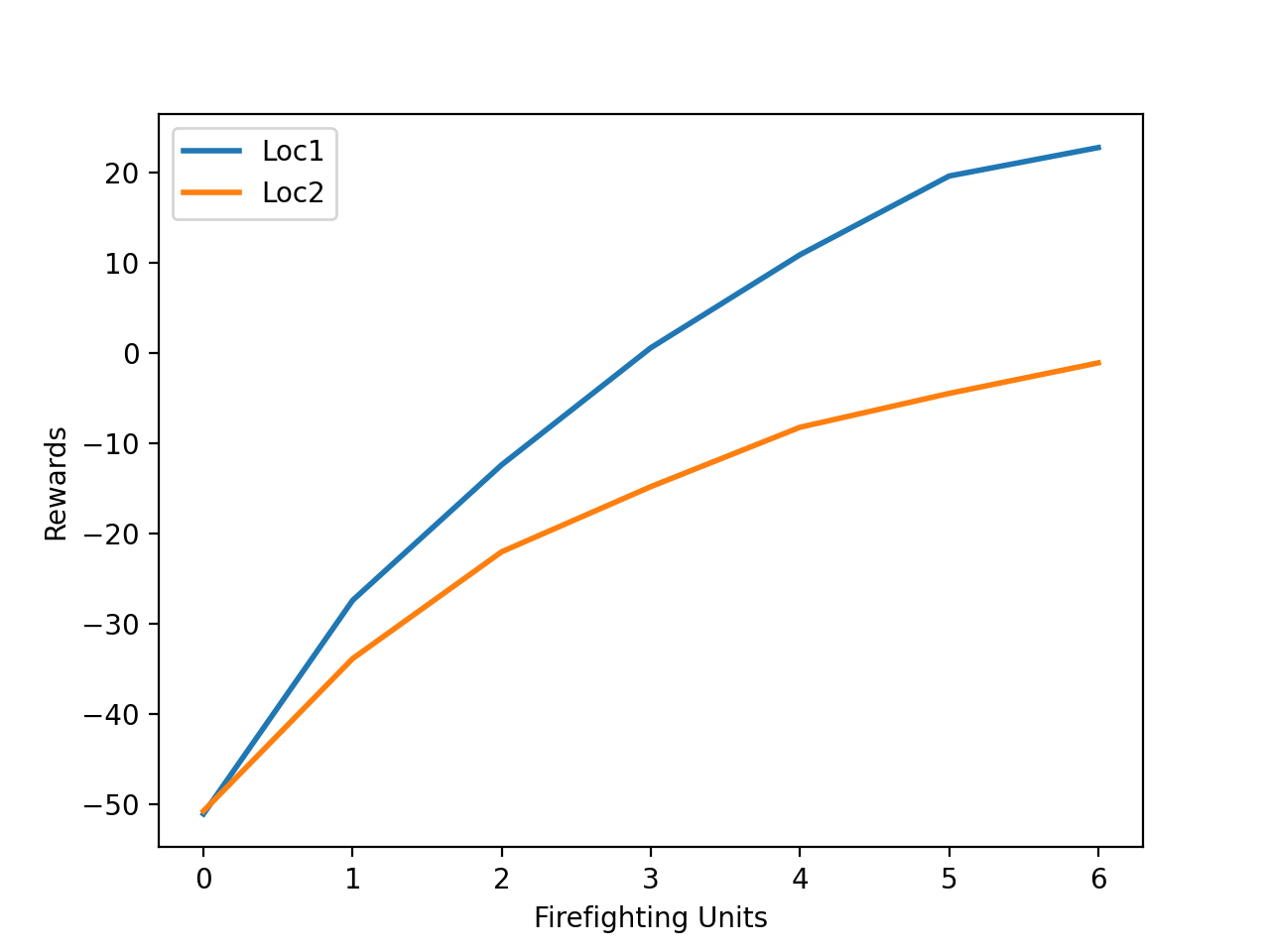}
    \caption{Total utility vs allocated firefighting units to a location}
    \label{fig:rewvsfire}
\end{figure}

We use a multi-objective reward function to incentivize extinguishing cells that are on fire while penalizing the overall spread of fire at each time unit. The locations have a size of $16 \times 16$. We set the discount factor to $\gamma = 0.95$, and train the DRL model for 15 minutes on a single CPU. This simple training provides us with a fairly effective and intuitive firefighting behavior, which allows us to calculate $\hat{F}_\ell (\mathrm{y})$. An example of this is shown in Fig.~\ref{fig:rewvsfire}, where we observe a diminishing return in the total discounted reward with respect to allocated resources. 
The environmental conditions at Loc.~1, including stronger winds and a specific vegetation profile, make it more susceptible to fire spread. Consequently, we have observed higher levels of total utility in fire suppression efforts at this location, as the DRL model employed is able to more effectively identify and navigate routes that lead to the extinguishment of ablaze cells in a shorter time frame. Furthermore, deploying a greater number of agents at this location has a more substantial impact on fire suppression efforts, in comparison to Loc.~2.

\begin{figure}[tb]
    \centering
    \includegraphics[width = \linewidth]{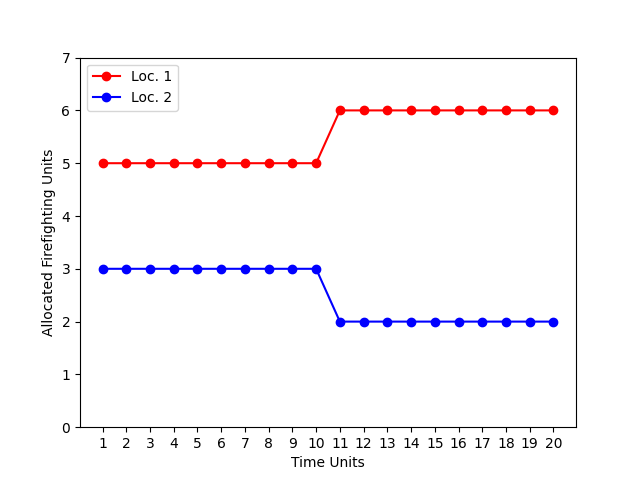}
    \caption{An example run of higher-layer allocations for the wildfire case with $z = 8$}
    \label{fig:FFvsTime}
\end{figure}

\begin{figure}[tb]
    \centering
    \includegraphics[width = \linewidth]{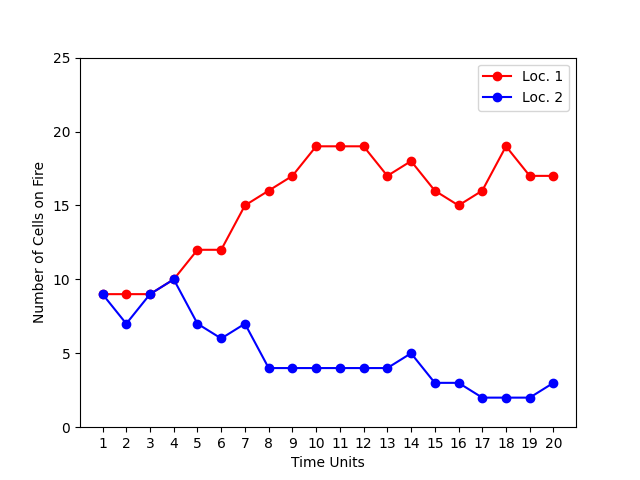}
    \caption{Number of cells on fire vs time-units accompanying the allocations in Fig.~\ref{fig:FFvsTime}}
    \label{fig:CellvsTime}
\end{figure}

As the next step, we integrate the results from the local systems to determine higher-level allocations in a rolling window manner, similar to the approach used in pandemic response. This integration is carried out using the system outlined in Eqns. (\ref{yIter})-(\ref{lambdaIter}). An example of higher-layer allocation process can be seen in Figs. (\ref{fig:FFvsTime}) and (\ref{fig:CellvsTime}). 
In this exemplary system, the look-ahead horizon is set to 24 and the allocation update period is every 10 time units. The allocations are based on both the actual fire spread and predictions, resulting in unique outcomes for each run. In this specific run, it is observed that the fire spreads significantly more in the first location, leading to the reallocation of one unit from Loc.~2 to Loc.~1. 
Clearly, this example is just a simple exposition, and in a real life implementation, there would be significant practical considerations to take into account. Nevertheless, it illustrates the theoretical approach of our study.

\section{Conclusions}
We present a methodology for addressing the complex problem of stochastic dynamic network utility maximization in the context of resource allocation, with a specific emphasis on the domain of disaster management. By taking into account the increasingly prevalent reality of the heterogeneous and hierarchical nature of large-scale incident response, the proposed approach utilizes a divide-and-conquer strategy through the implementation of a primal-dual framework to decompose the problem into more manageable localities. The local allocation of resources to individual entities is then addressed through the utilization of advanced deep reinforcement learning algorithms, wherein agent-based simulations are employed to model the underlying dynamics of the disaster scenario. These locally-derived solutions are informed by both real-time data from the ground as well as predictive forecasts, and the proposed methodology also incorporates a market mechanism for higher-level resource allocations. As such, the proposed approach does not necessitate a detailed understanding of the internal workings of the local systems at the upper level, with local entities effectively bidding for resources while the upper level dynamically sets pricing.

In this study, a comprehensive examination of the proposed methodology is presented, including a thorough exposition of the underlying theoretical foundations. The effectiveness of the method is subsequently demonstrated through the case studies of two distinct scenarios, specifically pandemic response and wildfire response. The results of the study provide insight into the potential utility of the proposed methodology in real-world applications.





 
\bibliographystyle{IEEEtran}
\bibliography{refs}
\end{document}